\documentclass[pra,amsmath,18pt]{revtex4}
\usepackage{mathrsfs}
\usepackage{graphicx}
\usepackage{epstopdf}
\usepackage{amsthm}
\usepackage{amsmath}
\usepackage{amssymb}
\usepackage{braket}
\usepackage{bbold}
\usepackage{subfigure}
\usepackage{hyperref}
\usepackage{appendix}
\usepackage{tikz}

\def\be{\begin{equation}}
\def\ee{\end{equation}}
\def\ba{\begin{array}}
\def\ea{\end{array}}

\def\1{{\bf{1}}}

\input amssym.def
\begin{document}
\title{Tight constraints of multi-qubit
entanglement in terms of nonconvex entanglement measures}

\smallskip
\author{Zhong-Xi Shen$^1$ }
\author{Dong-Ping Xuan$^1$ }
\author{Wen Zhou$^1$ }
\author{Zhi-Xi Wang$^1$ }
\author{Shao-Ming Fei$^{1,2}$}
\thanks{Corresponding author: feishm@cnu.edu.cn}
\affiliation{
$^1$School of Mathematical Sciences, Capital Normal University, Beijing 100048, China\\
$^2$Max-Planck-Institute for Mathematics in the Sciences, 04103 Leipzig, Germany
}

\begin{abstract}
In relation to nonconvex entanglement measures, we elucidate the constraints of multi-qubit entanglement, encompassing both the realms of monogamy and polygamy.  By using the Hamming weight of the binary vector related to the distribution of subsystems proposed in Kim (Sci. Rep. 8: 12245, 2018), we establish a class of monogamy inequalities for multi-qubit entanglement based on the $\alpha$th ($\alpha\geq 4\ln2$) power of logarithmic convex-roof extended negativity (LCREN), and a class of polygamy inequalities for multi-qubit entanglement in terms of the $\alpha$th ($0 \leq \alpha \leq 2$) power of logarithmic convex-roof extended negativity of assistance (LCRENoA).  For the case $\alpha<0$, we give the corresponding polygamy and monogamy relations for LCREN and LCRENoA, respectively. We also show that these new inequalities give rise to tighter constraints than the existing ones.  Moreover, our monogamy inequality is shown to be more effective for the counterexamples of the CKW monogamy inequality in higher-dimensional systems. Detailed examples are presented.\\

\medskip

\end{abstract}

\maketitle

\section{Introduction}\label{Intro}
Quantum  entanglement,  an  essential  aspect  of  quantum  mechanics,  provides  deep  understanding  of  the  nature  of  quantum  correlations  by  revealing  its  foundational  principles. One unique characteristic of quantum entanglement, which sets it apart from classical systems, is its limited shareability in multi-party quantum systems,
known as the  monogamy of entanglement (MoE)~\cite{T04, KGS}.  MoE is the fundamental ingredient for secure quantum cryptography~\cite{qkd1, qkd2}, and it also plays an important role in condensed-matter physics such as the $N$-representability problem for fermions~\cite{anti}.

Mathematically, MoE is characterized in a quantitative way known as the  monogamy inequality; for a three-qubit quantum state $\rho_{ABC}$
with its two-qubit reduced density matrices $\rho_{AB}=Tr_C \rho_{ABC}$ and
$\rho_{AC}=Tr_B \rho_{ABC}$, the first monogamy inequality was established by Coffman-Kundu-Wootters~(CKW) as
\begin{equation*}
\tau\left(\rho_{A|BC}\right)\geq \tau\left(\rho_{A|B}\right)+\tau\left(\rho_{A|C}\right)
\label{MoE}
\end{equation*}
where $\tau\left(\rho_{A|BC}\right)$ is the bipartite entanglement between subsystems $A$ and $BC$, quantified by  tangle and
$\tau\left(\rho_{A|B}\right)$ and $\tau\left(\rho_{A|C}\right)$ are the tangle between $A$ and $B$
and between $A$ and $C$, respectively~\cite{CKW}.

The CKW inequality demonstrates the mutually exclusive relationship of two-qubit entanglement
between $A$ and each of $B$ and $C$ measured by $\tau\left(\rho_{A|B}\right)$ and $\tau\left(\rho_{A|C}\right)$ respectively. As a result, the sum of the entanglement of the two-qubit systems cannot exceed the total entanglement between $A$ and $BC$, that is, $\tau\left(\rho_{A|BC}\right)$. Subsequently, the CKW inequality was generalized for arbitrary multi-qubit systems~\cite{OV} and extended to encompass multi-party and higher-dimensional quantum systems beyond qubits in some certain cases  in terms of various bipartite entanglement measures~\cite{KDS, KSRenyi, KimT, KSU}.

Whereas entanglement monogamy characterizes the restricted ability to share entanglement in multi-qubit quantum systems, the  assisted entanglement, which is a dual amount to bipartite entanglement measures, is also known to be
dually monogamous, thus polygamous in multi-qubit quantum systems;
for a three-qubit state $\rho_{ABC}$, a  polygamy inequality was proposed as
\begin{equation*}
\tau^a\left(\rho_{A|BC}\right)\leq\tau^a\left(\rho_{A|B}\right)
+\tau^a\left(\rho_{A|C}\right),
\label{PoE}
\end{equation*}
where $\tau^a\left(\rho_{A|BC}\right)$ is the tangle of assistance~\cite{GMS, GBS}. Later, the tangle-based polygamy inequality of entanglement was generalized into multi-qubit systems as well as some class of higher-dimensional quantum systems using various entropic entanglement measures~\cite{KimT, BGK, KUP}. General polygamy inequalities of entanglement were also formulated for multi-qubit quantum systems in arbitrary dimensions.~\cite{KimGP, KimGP16}.

Recently, a new monogamy inequalities employing entanglement measures raised to the power of $\alpha$ were proposed;
it was shown that the $\alpha$th-powered of  entanglement of formation and concurrence can be used to establish multi-qubit monogamy inequalities for $\alpha \geq \sqrt{2}$ and $\alpha \geq 2$, respectively~\cite{Fei}.
Later, tighter monogamy and polygamy inequalities of entanglement using non-negative power of concurrence and squar of convex-roof extended negativity were also proposed for multi-qubit systems~\cite{Fei2, Kim18}.

It is widely recognized that entanglement measures with convexity always satisfy monogamy inequalities.
Gao \emph{et.al} in Ref.~\cite{LMG} present a measure of entanglement, logarithmic convex-roof extended negativity (LCREN) satisfying important characteristics of an entanglement measure, and investigate the monogamy relation for  logarithmic negativity and LCREN both without convexity. They show exactly that the $\alpha$th power of logarithmic negativity, and a newly defined good measure of entanglement, LCREN, obey a class of general monogamy inequalities in $2\otimes2\otimes3$ systems and $2\otimes2\otimes2^{n}$ systems and multi-qubit systems for $\alpha\geq4\ln2$. They also  provide a class of general polygamy inequalities of multi-qubit systems in terms of logarithmic convex-roof extended negativity of assistance (LCRENoA) for $0\leq\alpha\leq2$.

In this paper, we provide a finer characterization of multi-qubit entanglement in
terms of nonconvex entanglement measures. By using the Hamming weight of the binary vectors related to
the subsystems, we establish a class of monogamy inequalities for multiqubit entanglement based on the $\alpha$th power of LCREN for $\alpha\geq4\ln2$. For $0\leq\alpha\leq2$, we establish a class of polygamy inequalities for multi-qubit entanglement in terms of the $\alpha$th power of LCRENoA. Even for the case of
$\alpha<0$, we can also provide tight constraints in terms of LCREN and LCRENoA. Thus, a complete
characterization for the full range of the power $\alpha$ is given. We further show that our class of monogamy and polygamy inequalities hold in a tighter way than those provided before~\cite{LMG}.  Moreover, our
monogamy inequality is shown to be more effective for the counterexamples of the CKW monogamy inequality in higher-dimensional systems.

\section{Preliminaries}\label{Pre}
We first recall the conceptions of  LCREN and  LCRENoA, and multi-qubit monogamy and polygamy inequalities.
For a quantum state $\rho_{AB}$ on Hilbert space $\mathcal{H}_{A}\otimes \mathcal{H}_{B}$, its negativity, $\mathcal{N}(\rho_{AB})$ is defined as \cite{GRF,MBP,KPAM}
\begin{equation}\label{2}
\mathcal{N}(\rho_{AB})=\|\rho_{AB}^{T_A}\|_{1}-1,
\end{equation}
where $\rho_{AB}^{T_A}$ denotes the partial transpose of $\rho_{AB}$ with respect to the subsystem $A$, and the trace norm $\|X\|_{1}=\text{tr}\sqrt{X X^\dag}$.

A more easily interpreted and computable measure of entanglement is the logarithmic negativity, which is defined as \cite{GRF, MBP}
\begin{equation}\label{4}
E_{\mathcal{N}}(\rho_{AB})=\log_{2}\|\rho_{AB}^{T_A}\|_{1}=\log_{2}[\mathcal{N}(\rho_{AB})+1].
\end{equation}
This quantity is an entanglement monotone both under general LOCC and PPT preserving operations but not convex \cite{MBP}. It is, moreover, additive.

Due to its construction, the negativity does not recognize entanglement in PPT states. In order to overcome its lack of separability criterion, one modification of negativity  is convex-roof extended negativity (CREN), which gives a perfect discrimination of PPT bound entangled states and separable states in any bipartite quantum system.

For a bipartite  state $\rho_{AB}$, its CREN, $\mathcal{\widetilde{N}}(\rho_{AB})$, is defined by  \cite{AAC}
\begin{equation}\label{5}
\mathcal{\widetilde{N}}(\rho_{AB})=\min_{\{p_{k}, |\varphi_{k}\rangle_{AB}\}}\sum_{k}{p_{k}}{\mathcal{N}}(|\varphi_{k}\rangle_{AB}),
\end{equation}
while the CREN of assistance (CRENoA), which can be considered
to be dual to CREN,  is defined  as \cite{KDS}
\begin{equation}\label{6}
\mathcal{\widetilde{N}}_a(\rho_{AB})=\max_{\{p_{k}, |\varphi_{k}\rangle_{AB}\}}\sum_{k}{p_{k}}{\mathcal{N}}(|\varphi_{k}\rangle_{AB}),
\end{equation}
where the minimum and maximum are taken over all possible pure-state decompositions of $\rho_{AB}=\sum_{k}p_{k}|\varphi_{k}\rangle_{AB}\langle\varphi_{k}|$.
By definition, both the CREN and CRENoA of a pure state are equal to its negativity.

For any bipartite state $\rho_{AB}$, we define LCREN as
\begin{equation}\label{9}
E_{\mathcal{\widetilde{N}}}(\rho_{AB})=\log_{2}[\mathcal{\widetilde{N}}(\rho_{AB})+1].
\end{equation}

Clearly, LCREN is invariant under local unitary transformations. One important property is this: $E_{\mathcal{\widetilde{N}}}(\rho_{AB})$ is nonzero if
and only if $\rho_{AB}$ is entangled (and so it equals zero if
and only if $\rho_{AB}$ is separable). Besides, it is entanglement monotone under LOCC operations. LCREN is not only nonincreasing under LOCC, but also nonincreasing on average under
LOCC, which follow from the entanglement monotonicity of CREN under LOCC, the monotonicity logarithm, and concavity of logarithm.

However, just as logarithmic negativity, LCREN is also not convex. Suppose that  $\rho_{AB}=\sum_{k}p_{k}\rho_k$ with $\rho_k=|\varphi_k\rangle_{AB}\langle\varphi_k|$ is the optimal decomposition for $\rho_{AB}$
achieving the minimum of (\ref{5}). Then  $\mathcal{\widetilde{N}}(\rho_{AB})= \sum_{k}{p_{k}}{\mathcal{N}}(|\varphi_{k}\rangle_{AB})$ by definition.  The concavity of logarithm ensures
\begin{equation}
\begin{array}{cl}
  E_{\mathcal{\widetilde{N}}}(\sum_{k}p_{k}\rho_k) & =\log_2[\sum_{k}{p_{k}}{\mathcal{N}}(|\varphi_{k}\rangle_{AB})+1] \\
  & =\log_2[\sum_{k}{p_{k}}\|\rho_k^{T_A}\|_1] \\
   & \geq \sum_{k}{p_{k}} \log_2\|\rho_k^{T_A}\|_1\\
  & = \sum_{k}p_{k} E_{\mathcal{\widetilde{N}}}(\rho_k),
\end{array}
\end{equation}
which implies that LCREN is not convex.

We can also show by concrete examples that it is not convex. Consider the mixed qubit state $\rho=\frac{1}{2}(\rho_{1}+\rho_{2})$ with $\rho_{1}=\frac{|01\rangle-|10\rangle}{\sqrt{2}}\frac{\langle01|-\langle10|}{\sqrt{2}}$ and $\rho_{2}=|01\rangle\langle01|$. By definition of $\mathcal{\widetilde{N}}(\rho_{AB})$ (Eq.(\ref{5})), for two qubit state $\rho_{AB}$, we have
\begin{equation}
\mathcal{\widetilde{N}}(\rho_{AB})=\min_{\{p_{k}, |\varphi_{k}\rangle_{AB}\}}\sum_{k}{p_{k}}{\mathcal{N}}(|\varphi_{k}\rangle_{AB})=\mathcal{C}(\rho_{AB}),
\end{equation}
because the negativity is the concurrence for the pure states. Here
$\mathcal{C}(\rho_{AB})$ is the concurrence of the mixed qubit state~\cite{Woo}. According
to Ref.~\cite{Woo} we can obtain
$$\mathcal{\widetilde{N}}(\rho)=\frac{1}{2},\quad\mathcal{\widetilde{N}}(\rho_{1})=1,\quad\mathcal{\widetilde{N}}(\rho_{2})=0.$$
So one has
$$E_{\mathcal{\widetilde{N}}}(\rho)=\log_{2}\frac{3}{2},\quad E_{\mathcal{\widetilde{N}}}(\rho_{1})=1,\quad E_{\mathcal{\widetilde{N}}}(\rho_{2})=0,$$
from which it easily follows that
$$E_{\mathcal{\widetilde{N}}}(\rho)>\frac{1}{2}E_{\mathcal{\widetilde{N}}}(\rho_{1})+\frac{1}{2}E_{\mathcal{\widetilde{N}}}(\rho_{2}).$$
This implies that LCREN is not convex.

For any multiqubit state $\rho_{A B_0 \cdots B_{N-1}}$,
a monogamous inequality has been presented in Ref.~\cite{LMG} for $\alpha \geq 4\ln2$,
\begin{equation}\label{mono}
E_{\mathcal{\widetilde{N}}}^{\alpha}(\rho_{A|B_{0}\cdots B_{N-1}})
 \geq \sum\limits_{i=0}^{N-1}  E_{\mathcal{\widetilde{N}}}^{\alpha}(\rho_{A|B_{i}}),
\end{equation}
where $E_{\mathcal{\widetilde{N}}}(\rho_{A|B_{0}\cdots B_{N-1}})$ is the
LCREN of $\rho_{A B_0 \cdots B_N-1}$ with respect to the bipartition between $A$ and $B_0 \cdots B_{N-1}$, and $E_{\mathcal{\widetilde{N}}}(\rho_{A|B_{i}})$ is the LCREN of the reduced density matrix
$\rho_{A B_i}$, $i=0,\cdots,N-1$.

Similar to the duality between CREN and CRENoA, we can also define a dual to LCREN, namely LCRENoA, by
\begin{equation}\label{10}
E_{\mathcal{\widetilde{N}}_a}(\rho_{AB})=\log_{2}[\mathcal{\widetilde{N}}_a(\rho_{AB})+1].
\end{equation}

In addition, a class of polygamy inequalities has been obtained for multi-qubit systems in Ref.~\cite{LMG},
\begin{equation}\label{poly}
E_{\mathcal{\widetilde{N}}_a}^{\alpha}(\rho_{A|B_{0}\cdots B_{N-1}})
 \leq \sum\limits_{i=0}^{N-1} E_{\mathcal{\widetilde{N}}_a}^{\alpha}(\rho_{A|B_{i}}),
\end{equation}
for $0 \leq \alpha \leq 2$, $\alpha \neq 1$, where $E_{\mathcal{\widetilde{N}}_a}(\rho_{A|B_{0}\cdots B_{N-1}})$ is the LCRENoA of $\rho_{A B_0 \cdots B_N-1}$ with
respect to the bipartition between $A$ and $B_0 \cdots B_{N-1}$, and $E_{\mathcal{\widetilde{N}}_a}(\rho_{A|B_{i}})$ is the LCRENoA of the reduced density matrix $\rho_{AB_i}$, $i=0,\cdots,N-1$.

In the following we show that these inequalities above can be further improved to be much tighter under certain conditions, which provide tighter constraints on the multiqubit entanglement distribution.

\section{Tight constraints of multi-qubit entanglement in terms of LCREN}
Here we establish a class of tight monogamy inequalities of multi-qubit entanglement using the $\alpha$th-powered of LCREN. Before we present our main results, we first provide some notations, definitions and a lemma, which are useful throughout this paper.

In Ref.~\cite{SCI}, Kim established a class of tight monogamy inequalities of multiqubit entanglement in terms of Hamming weight. For any nonnegative integer $j$ with binary expansion
$j=\sum_{i=0}^{n-1} j_i 2^i$, where $\log_{2}j \leq n$ and $j_i \in \{0, 1\}$ for $i=0, \cdots, n-1$,
one can always define a unique binary vector associated with $j$,
$\overrightarrow{j}=\left(j_0,~j_1,~\cdots ,j_{n-1}\right)$.
The Hamming weight $\omega_{H}\left(\overrightarrow{j}\right)$ of the binary vector $\overrightarrow{j}$
is defined to be the number of $1's$ in its coordinates ~\cite{nc}.
Moreover, the Hamming weight $\omega_{H}\left(\overrightarrow{j}\right)$ is bounded above by $\log_{2}j$,
\begin{equation}\label{n11}
\omega_{H}\left(\overrightarrow{j}\right)\leq \log_{2}j \leq j.
\end{equation}
We also provide the following lemma whose proof is easily obtained by some straightforward calculus.

\noindent{[\bf Lemma 1]}.
For $x \in \left[0,1\right]$ and nonnegative real numbers $\alpha, \beta$, we have
\begin{align}\label{bu1}
\left(1+x\right)^{\alpha}\geq1+\alpha x^{\alpha}
\end{align}
for $\alpha \geq1$, and
\begin{align}\label{bu2}
\left(1+x\right)^{\alpha}\leq 1+\alpha x^{\alpha}
\end{align}
for $0\leq\alpha \leq1$.

Now we provide our first result, which states that a class of tight monogamy inequalities of multi-qubit entanglement can be established using the $\alpha$th-powered LCREN and the Hamming weight of the binary vector related with the distribution of subsystems.

\noindent{[\bf Theorem 1]}. For any multi-qubit state $\rho_{AB_0\ldots B_{N-1}}$ and $\alpha\geq 4\ln2$, we have
\begin{equation}\label{n18}
[E_{\mathcal{\widetilde{N}}}(\rho_{A|B_0B_1\ldots B_{N-1}})]^\alpha
\geq\sum\limits_{j=0}^{N-1}\Big(\frac{\alpha}{4\ln2}\Big)^{\omega_H(\vec{j})}[E_{\mathcal{\widetilde{N}}}(\rho_{A|B_j})]^\alpha,
\end{equation}
where $\alpha\geq4\ln2$, $\overrightarrow{j}=\left(j_0, \cdots ,j_{n-1}\right)$ is the vector from the binary representation of $j$, and
$\omega_{H}\left(\overrightarrow{j}\right)$ is the Hamming weight of $\overrightarrow{j}$.

\noindent{[\bf Proof]}. From inequality \ref{mono},  one has $E_{\mathcal{\widetilde{N}}}^{4\ln2}(\rho_{A|B_{0}\cdots B_{N-1}})\geq \sum\limits_{i=0}^{N-1}  E_{\mathcal{\widetilde{N}}}^{4\ln2}(\rho_{A|B_{i}})$, thus, it is sufficient to show that
\begin{equation}\label{n19}
\left[\sum\limits_{j=0}^{N-1} E_{\mathcal{\widetilde{N}}}^{4\ln2}(\rho_{A|B_j})\right]^\frac{\alpha}{4\ln2}
\geq\sum\limits_{j=0}^{N-1}\Big(\frac{\alpha}{4\ln2}\Big)^{\omega_{H}(\vec{j})}
[E_{\mathcal{\widetilde{N}}}(\rho_{A|B_j})]^\alpha.
\end{equation}
Without loss of generality, we assume that the qubit subsystems $B_0, \ldots, B_{N-1}$ are so labeled such that
\begin{equation}\label{n20}
E_{\mathcal{\widetilde{N}}}^{4\ln2}(\rho_{A|B_j})\geq E_{\mathcal{\widetilde{N}}}^{4\ln2}(\rho_{A|B_{j+1}})\geq 0
\end{equation}
for $j=0,1,\ldots,N-2$.

We first show that the inequality \eqref{n19} holds for the case of $N=2^n$.
For $n=1$, let $\rho_{AB_0}$ and $\rho_{AB_1}$ be the two-qubit reduced density matrices
of a three-qubit pure state $\rho_{AB_0 B_1}$. We obtain
\begin{equation}\label{n21}
[E_{\mathcal{\widetilde{N}}}^{4\ln2}(\rho_{A|B_0})+E_{\mathcal{\widetilde{N}}}^{4\ln2}(\rho_{A|B_1})]^\frac{\alpha}{4\ln2}
=[E_{\mathcal{\widetilde{N}}}(\rho_{A|B_0})]^\alpha \Big(1+\frac{E_{\mathcal{\widetilde{N}}}^{4\ln2}(\rho_{A|B_1})}{E_{\mathcal{\widetilde{N}}}^{4\ln2}(\rho_{A|B_0})}\Big)^\frac{\alpha}{4\ln2}.
\end{equation}
Combining \eqref{bu1} and \eqref{n20}, we have
\begin{equation}\label{n22}
\Big(1+\frac{E_{\mathcal{\widetilde{N}}}^{4\ln2}(\rho_{A|B_1})}{E_{\mathcal{\widetilde{N}}}^{4\ln2}(\rho_{A|B_0})}\Big)^\frac{\alpha}{4\ln2} \geq
1+\displaystyle\frac{\alpha}{4\ln2}\Bigg(\frac{E_{\mathcal{\widetilde{N}}}(\rho_{A|B_1})}{E_{\mathcal{\widetilde{N}}}(\rho_{A|B_0})}\Bigg)^\alpha.
\end{equation}
From \eqref{n21} and \eqref{n22}, we get
\begin{equation*}
[E_{\mathcal{\widetilde{N}}}^{4\ln2}(\rho_{A|B_0})+E_{\mathcal{\widetilde{N}}}^{4\ln2}(\rho_{A|B_1})]^\frac{\alpha}{4\ln2}\geq
[E_{\mathcal{\widetilde{N}}}(\rho_{A|B_0})]^\alpha+\displaystyle\frac{\alpha}{4\ln2}[E_{\mathcal{\widetilde{N}}}(\rho_{A|B_1})]^\alpha.
\end{equation*}
Therefore, the inequality \eqref{n19} holds for $n=1$.

We assume that the inequality \eqref{n19} holds for $N=2^{n-1}$ with $n\geq 2$,
and prove the case of $N=2^n$.
For an $(N+1)$-qubit state $\rho_{AB_0B_1 \cdots B_{N-1}}$ with its two-qubit reduced density matrices $\rho_{AB_j}$ with $j=0, \cdots, N-1$,
we have
\begin{align}\label{duo1}
\left(\sum_{j=0}^{N-1}E_{\mathcal{\widetilde{N}}}^{4\ln2}\left(\rho_{A|B_j}\right)\right)^\frac{\alpha}{4\ln2}=
\left(\sum_{j=0}^{2^{n-1}-1}E_{\mathcal{\widetilde{N}}}^{4\ln2}\left(\rho_{A|B_j}\right)\right)^\frac{\alpha}{4\ln2}
\left(1+\frac{\sum_{j=2^{n-1}}^{2^n-1}E_{\mathcal{\widetilde{N}}}^{4\ln2}\left(\rho_{A|B_j}\right)}
{\sum_{j=0}^{2^{n-1}-1}E_{\mathcal{\widetilde{N}}}^{4\ln2}\left(\rho_{A|B_j}\right)}\right)^\frac{\alpha}{4ln2}.
\end{align}
Because the ordering of subsystems in Inequality \ref{n20} implies
\begin{equation*}
0\leq\frac{\sum\nolimits_{j=2^{n-1}}^{2^n-1}E_{\mathcal{\widetilde{N}}}^{4\ln2}(\rho_{A|B_j})}{\sum\nolimits_{j=0}^{2^{n-1}-1}
E_{\mathcal{\widetilde{N}}}^{4\ln2}(\rho_{A|B_j})}\leq 1.
\end{equation*}
Thus, the Eq.~(\ref{duo1}) and Inequality~(\ref{bu1}) lead us to
\begin{equation*}
\left(\sum_{j=0}^{N-1}E_{\mathcal{\widetilde{N}}}^{4\ln2}\left(\rho_{A|B_j}\right)\right)^\frac{\alpha}{4\ln2}
\geq \Bigg(\sum\nolimits_{j=0}^{2^{n-1}-1}E_{\mathcal{\widetilde{N}}}^{4\ln2}(\rho_{A|B_j})\Bigg)^\frac{\alpha}{4\ln2}
+\displaystyle\frac{\alpha}{4\ln2}\Bigg(\sum\nolimits_{j=2^{n-1}}^{2^n-1}E_{\mathcal{\widetilde{N}}}^{4\ln2}(\rho_{A|B_j})\Bigg)^\frac{\alpha}{4\ln2}.
\end{equation*}
According to the induction hypothesis, we get
$$
\Bigg(\sum\nolimits_{j=0}^{2^{n-1}-1}E_{\mathcal{\widetilde{N}}}^{4\ln2}(\rho_{A|B_j})\Bigg)^\frac{\alpha}{4\ln2} \geq
\sum\nolimits_{j=0}^{2^{n-1}-1}\Big(\frac{\alpha}{4\ln2}\Big)^{\omega_H(\vec{j})}
[E_{\mathcal{\widetilde{N}}}(\rho_{A|B_j})]^\alpha.
$$
By relabeling the subsystems, the induction hypothesis leads to
$$
\Bigg(\sum\nolimits_{j=2^{n-1}}^{2^n-1}E_{\mathcal{\widetilde{N}}}^{4\ln2}(\rho_{A|B_j})\Bigg)^\frac{\alpha}{4\ln2} \geq
\sum\nolimits_{j=2^{n-1}}^{2^n-1}\Big(\frac{\alpha}{4\ln2}\Big)^{\omega_H(\vec{j})-1}[E_{\mathcal{\widetilde{N}}}(\rho_{A|B_j})]^\alpha.
$$
Thus, we have
$$
\Bigg(\sum\nolimits_{j=0}^{2^n-1}E_{\mathcal{\widetilde{N}}}^{4\ln2}(\rho_{A|B_j})\Bigg)^\frac{\alpha}{4\ln2}\geq
\sum\nolimits_{j=0}^{2^n-1}\Big(\frac{\alpha}{4\ln2}\Big)^{\omega_H(\vec{j})}[E_{\mathcal{\widetilde{N}}}(\rho_{A|B_j})]^\alpha.
$$
Now consider a $(2^n+1)$-qubit state
\begin{equation}\label{n27}
\Gamma_{AB_0B_1\ldots B_{2^n-1}}=\rho_{AB_0B_1\ldots B_{N-1}}\otimes \sigma_{B_N\ldots B_{2^n-1}},
\end{equation}
which is the tensor product of $\rho_{AB_0B_1\ldots B_{N-1}}$ and an arbitrary $(2^n-N)$-qubit state $\sigma_{B_N\ldots B_{2^n-1}}$. We have
\begin{equation*}
[E_{\mathcal{\widetilde{N}}}^{4\ln2}(\Gamma_{A|B_0B_1\ldots B_{2^n-1}})]^\frac{\alpha}{4\ln2}
\geq\sum\nolimits_{j=0}^{2^n-1}\Big(\frac{\alpha}{4\ln2}\Big)^{\omega_H(\vec{j})}[E_{\mathcal{\widetilde{N}}}(\Gamma_{A|B_j})]^\alpha,
\end{equation*}
where $\Gamma_{A|B_j}$ is the two-qubit reduced density matrix of $\Gamma_{AB_0B_1\ldots B_{2^n-1}}$, $j=0,1,\ldots,2^n-1$. Therefore,
\begin{align*}
[E_{\mathcal{\widetilde{N}}}^{4\ln2}(\rho_{A|B_0B_1\ldots B_{N-1}})]^\frac{\alpha}{4\ln2} =&[E_{\mathcal{\widetilde{N}}}^{4\ln2}(\Gamma_{A|B_0B_1\ldots B_{2^n-1}})]^\frac{\alpha}{4\ln2}\nonumber\\
\geq& \sum\nolimits_{j=0}^{2^n-1}\Big(\frac{\alpha}{4\ln2}\Big)^{\omega_H(\vec{j})}[E_{\mathcal{\widetilde{N}}}(\Gamma_{A|B_j})]^\alpha \nonumber\\
=& \sum\nolimits_{j=0}^{N-1}\Big(\frac{\alpha}{4\ln2}\Big)^{\omega_H(\vec{j})}[E_{\mathcal{\widetilde{N}}}(\rho_{A|B_j})]^\alpha,
\end{align*}
where $\Gamma_{A|B_0B_1\ldots B_{2^n-1}}$ is separated to the bipartition $AB_0\ldots B_{N-1}$ and $B_N\ldots B_{2^n-1}$, $E_{\mathcal{\widetilde{N}}}\left(\Gamma_{A|B_0 B_1 \cdots B_{2^n-1}}\right)=E_{\mathcal{\widetilde{N}}}\left(\rho_{A|B_0 B_1 \cdots B_{N-1}}\right)$,
$E_{\mathcal{\widetilde{N}}}\left(\Gamma_{A|B_j}\right)=0$ for $j=N, \cdots , 2^n-1$,
and $\Gamma_{AB_j}=\rho_{AB_j}$ for each $j=0, \cdots , N-1$, and this completes the proof.
\qed

\noindent{[\bf Remark 1]}. Since $(\frac{\alpha}{4\ln2})^{\omega_H(\overrightarrow{j})}\geqslant 1$ for  any $\alpha\geq4\ln2$, for any multi-qubit state $\rho_{AB_0 B_1 \cdots B_{N-1}}$ we have the following relation
\begin{equation*}
[E_{\mathcal{\widetilde{N}}}(\rho_{A|B_0B_1\ldots B_{N-1}})]^\alpha
\geq\sum\nolimits_{j=0}^{N-1}\Big(\frac{\alpha}{4\ln2}\Big)^{\omega_H(\vec{j})}[E_{\mathcal{\widetilde{N}}}(\rho_{A|B_j})]^\alpha
\geq\sum\nolimits_{j=0}^{N-1}[E_{\mathcal{\widetilde{N}}}(\rho_{A|B_j})]^\alpha.
\end{equation*}
Therefore, our inequality \eqref{n18} in Theorem 1 is always tighter than the inequality \eqref{mono} in Ref.~\cite{LMG}.

\begin{figure}[h]
	\centering
	\scalebox{2.0}{\includegraphics[width=3.9cm]{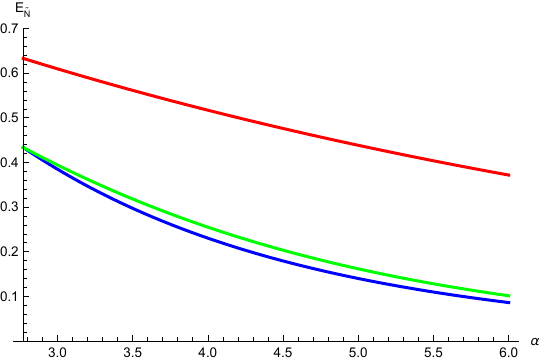}}
	\caption{\small The vertical axis is the the lower bound of the  LCREN $E_{\mathcal{\widetilde{N}}}(\rho_{A|BC})$. The red line is the exact values of  $E_{\mathcal{\widetilde{N}}}(\rho_{A|BC})$. The  green  line represents the lower bound from our results. The blue line represents the lower bound from the result in \cite{LMG}.}
	\label{Fig1}
\end{figure}

\noindent{[\bf Example 1]}.
Let us consider the three-qubit state $|\phi\rangle_{ABC}$ in the generalized Schmidt decomposition\cite{AAC,XHS},
\begin{equation}\label{GSD}
|\phi\rangle_{ABC}=\lambda_0|000\rangle+\lambda_1e^{i\varphi}|100\rangle+\lambda_2|101\rangle+\lambda_3|110\rangle+\lambda_4|111\rangle,
\end{equation}
where $\lambda_i\geq0$, $i=0,1,\cdots,4$, and $\sum\limits_{i=0}^{4}\lambda_i^2=1$.
One gets $\mathcal{\widetilde{N}}(\rho_{A|BC})=2\lambda_0\sqrt{\lambda_2^2+\lambda_3^2+\lambda_4^2}$, $\mathcal{\widetilde{N}}(\rho_{A|B})=2\lambda_0\lambda_2$ and $\mathcal{\widetilde{N}}(\rho_{A|C})=2\lambda_0\lambda_3$. Setting $\lambda_0=\lambda_3=\lambda_4={1}/{\sqrt{5}}$, $\lambda_2=\sqrt{{2}/{5}}$ and $\lambda_1=0$, we have $\mathcal{\widetilde{N}}(\rho_{A|BC})={4}/{5}$, $\mathcal{\widetilde{N}}(\rho_{A|B})={2\sqrt{2}}/{5}$ and $\mathcal{\widetilde{N}}(\rho_{A|C})={2}/{5}$. Using (\ref{9}) we have $E_{\mathcal{\widetilde{N}}}(\rho_{A|BC})=\log_{2}\frac{9}{5}$, $E_{\mathcal{\widetilde{N}}}(\rho_{A|B})=\log_{2}({2\sqrt{2}}/{5}+1)$ and $E_{\mathcal{\widetilde{N}}}(\rho_{A|C})=\log_{2}\frac{7}{5}$.
Thus, $[E_{\mathcal{\widetilde{N}}}(\rho_{A|BC})]^\alpha\geq(\log_{2}({2\sqrt{2}}/{5}+1))^\alpha
+\frac{\alpha}{4\ln2}(\log_{2}\frac{7}{5})^\alpha$ from our result (\ref{n18}), and  $[E_{\mathcal{\widetilde{N}}}(\rho_{A|BC})]^\alpha\geq(\log_{2}({2\sqrt{2}}/{5}+1))^\alpha
+(\log_{2}\frac{7}{5})^\alpha$ from the result given in Ref.~\cite{LMG}. One can see that our result is better than the result in Ref.~\cite{LMG} for $\alpha\geq4\ln2$, see Fig. \ref{Fig1}.

Under certain conditions, the inequality \eqref{n18} can even be improved
further to become a much tighter inequality.

\noindent{[\bf Theorem 2]}.
For $\alpha\geq4\ln2$, any multiqubit state $\rho_{AB_0\ldots B_{N-1}}$ satisfies
\begin{equation}\label{n31}
[E_{\mathcal{\widetilde{N}}}(\rho_{A|B_0B_1\ldots B_{N-1}})]^\alpha
\geq\sum\nolimits_{j=0}^{N-1}\Big(\frac{\alpha}{4\ln2}\Big)^j[E_{\mathcal{\widetilde{N}}}(\rho_{A|B_{j}})]^{\alpha},
\end{equation}
if
\begin{equation}\label{n32}
E_{\mathcal{\widetilde{N}}}^{4\ln2}(\rho_{A|B_i})\geq\sum\nolimits_{j=i+1}^{N-1}E_{\mathcal{\widetilde{N}}}^{4\ln2}(\rho_{A|B_j})
\end{equation}
for $i=0,1,\ldots,N-2$.

\noindent{[\bf Proof]}.
From inequality (\ref{mono}),  we only need to prove
	\begin{equation}\label{thm11:3}
		\left[\sum\limits_{j=0}^{N-1} E_{\mathcal{\widetilde{N}}}^{4\ln2}(\rho_{A|B_j})\right]^\frac{\alpha}{4\ln2}
\geq\sum\limits_{j=0}^{N-1}\Big(\frac{\alpha}{4\ln2}\Big)^{j}
[E_{\mathcal{\widetilde{N}}}(\rho_{A|B_j})]^\alpha.
	\end{equation}
	We use mathematical induction on $N$ here. It is obvious that inequality (\ref{thm11:3}) holds for $N=2$ from (\ref{n18}). Assume that it also holds for any positive integer less than $N$. Since $\frac{\sum\nolimits_{j=i+1}^{N-1}E_{\mathcal{\widetilde{N}}}^{4\ln2}(\rho_{A|B_j})}{E_{\mathcal{\widetilde{N}}}^{4\ln2}(\rho_{A|B_i})}\leqslant 1$, we have
	\begin{eqnarray}
		\left[\sum\limits_{j=0}^{N-1} E_{\mathcal{\widetilde{N}}}^{4\ln2}(\rho_{A|B_j})\right]^\frac{\alpha}{4\ln2}
		&=&E_{\mathcal{\widetilde{N}}}^{\alpha}(\rho_{A|B_0})		\Bigg(1+\frac{\sum\nolimits_{j=1}^{N-1}E_{\mathcal{\widetilde{N}}}^{4\ln2}(\rho_{A|B_j})}{E_{\mathcal{\widetilde{N}}}^{4\ln2}(\rho_{A|B_0})} \Bigg)^{\frac{\alpha}{4\ln2}}\nonumber\\
		&\geqslant& E_{\mathcal{\widetilde{N}}}^{\alpha}(\rho_{A|B_0})\Bigg[1+
\frac{\alpha}{4\ln2}\Bigg(\frac{\sum\nolimits_{j=1}^{N-1}E_{\mathcal{\widetilde{N}}}^{4\ln2}(\rho_{A|B_j})}{E_{\mathcal{\widetilde{N}}}^{4\ln2}(\rho_{A|B_0})}\Bigg)^{\frac{\alpha}{4\ln2}}\Bigg]\nonumber\\		&=&E_{\mathcal{\widetilde{N}}}^{\alpha}(\rho_{A|B_0})+\frac{\alpha}{4\ln2}\Bigg(\sum\nolimits_{j=1}^{N-1}E_{\mathcal{\widetilde{N}}}^{4\ln2}(\rho_{A|B_j})\Bigg)^{\frac{\alpha}{4\ln2}}\nonumber\\	&\geqslant& E_{\mathcal{\widetilde{N}}}^{\alpha}(\rho_{A|B_0})+\frac{\alpha}{4\ln2} \sum\limits_{j=1}^{N-1}(\frac{\alpha}{4\ln2})^{j-1}E_{\mathcal{\widetilde{N}}}^{\alpha}(\rho_{A|B_j})\nonumber\\
		&=&\sum\nolimits_{j=0}^{N-1}\Big(\frac{\alpha}{4\ln2}\Big)^j[E_{\mathcal{\widetilde{N}}}(\rho_{A|B_{j}})]^{\alpha}\nonumber,
	\end{eqnarray}
	where the first inequality is due to Lemma \ref{bu1}  and the second inequality is due to the induction hypothesis.
\qed

\noindent{[\bf Remark 2]}. In fact, according to \eqref {n11}, for any $\alpha \geq 4\ln2$, one has
\begin{align*}
[E_{\mathcal{\widetilde{N}}}(\rho_{A|B_0\ldots B_{N-1}})]^\alpha
\geq & \sum_{j=0}^{N-1} \left( \frac{\alpha}{4\ln2}\right)^{j}\left(E_{\mathcal{\widetilde{N}}}\left(\rho_{A|B_j}\right)\right)^{\alpha} \nonumber\\
\geq & \sum\nolimits_{j=0}^{N-1}\Big(\frac{\alpha}{4\ln2}\Big)^{\omega_H(\vec{j})}[E_{\mathcal{\widetilde{N}}}(\rho_{A|B_j})]^\alpha.
\end{align*}
Therefore the inequality (\ref{n31}) of Theorem 2 is tighter than  the inequality (\ref{n18}) of Theorem 1 under certain conditions.

\begin{figure}[h]
	\centering
	\scalebox{2.0}{\includegraphics[width=3.9cm]{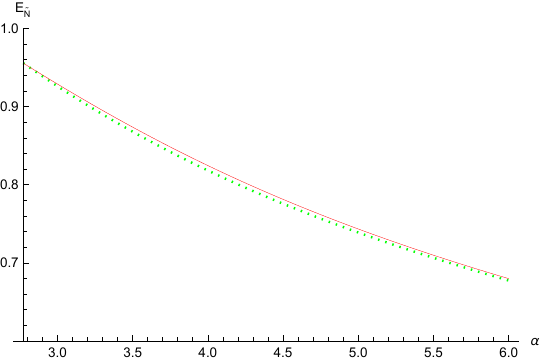}}
	\caption{\small The vertical axis is the the lower bound of the  LCREN $E_{\mathcal{\widetilde{N}}}(\rho_{A|BCD})$. The red  thin line represents the lower bound from our result (\ref{n31}). The green dotted line represents the lower bound from our result (\ref{n18}).}
	\label{Fig2}
\end{figure}

\noindent{[\bf Example 2]}. Let's consider a four-qubit entangled decoherence-free state is given by \cite{Zhou(2016)}:
\begin{eqnarray}
|\Phi\rangle=a|\Psi_0\rangle+b|\Psi_1\rangle,
\label{eqnfreestate}
\end{eqnarray}
where $|\Psi_i\rangle$ are logic basis states given by
\begin{eqnarray}
|\Psi_0\rangle_{ABCD}&=&\frac{1}{2}(|01\rangle-|10\rangle)_{AB}(|01\rangle-|10\rangle)_{CD},
\nonumber\\
|\Psi_1\rangle_{ABCD}&=&\frac{1}{2\sqrt{3}}(2|1100\rangle+
2|0011\rangle-|1010\rangle-|1001\rangle
\nonumber\\
&&-|0101\rangle-|0110\rangle)_{ABCD}.
\label{11}
\end{eqnarray}
When $a=b=\frac{1}{\sqrt{2}}$
the concurrence for $|\Phi\rangle$ are computed as
$\mathcal{\widetilde{N}}(|\Phi\rangle_{A|BCD})=1$,  $\mathcal{\widetilde{N}}(\rho_{A|B})=0.9107$,
$\mathcal{\widetilde{N}}(\rho_{A|C})=0.3333$ and $\mathcal{\widetilde{N}}(\rho_{A|D})=0.244$.
Using (\ref{9}) we have $E_{\mathcal{\widetilde{N}}}(\rho_{A|BCD})=1$, $E_{\mathcal{\widetilde{N}}}(\rho_{AB})=0.934101$, $E_{\mathcal{\widetilde{N}}}(\rho_{AC})=0.415001$
and $E_{\mathcal{\widetilde{N}}}(\rho_{AD})=0.314986$.
Thus, $[E_{\mathcal{\widetilde{N}}}(\rho_{A|BCD})]^\alpha\geq(0.934101)^\alpha
+\frac{\alpha}{4\ln2}(0.415001)^\alpha+(\frac{\alpha}{4\ln2})^{2}(0.314986)^\alpha$ from our result (\ref{n31}), and  $[E_{\mathcal{\widetilde{N}}}(\rho_{A|BCD})]^\alpha\geq(0.934101)^\alpha
+\frac{\alpha}{4\ln2}(0.415001)^\alpha+\frac{\alpha}{4\ln2}(0.314986)^\alpha$ from our result (\ref{n18}). One can see that our result (\ref{n31}) is better than the result (\ref{n18}) for $\alpha\geq4\ln2$, see Fig. \ref{Fig2}.

In general, the conditions (\ref{n32}) is not always satisfied. We derive the following monogamy inequality with different conditions.

\noindent{[\bf Theorem 3]}.
For $\alpha\geq4\ln2$, any multiqubit state $\rho_{AB_0\ldots B_{N-1}}$ satisfies
\begin{eqnarray}\label{ger1}
		[E_{\mathcal{\widetilde{N}}}(\rho_{A|B_0\ldots B_{N-1}})]^\alpha&\geqslant& \sum_{j=0}^{t}(\frac{\alpha}{4\ln2})^{j}[E_{\mathcal{\widetilde{N}}}(\rho_{A|B_{j}})]^\alpha+(\frac{\alpha}{4\ln2} )^{t+2}\sum_{j=t+1}^{N-2}[E_{\mathcal{\widetilde{N}}}(\rho_{A|B_{j}})]^\alpha\nonumber\\
		&&\ \ \
		+(\frac{\alpha}{4\ln2})^{t+1}[E_{\mathcal{\widetilde{N}}}(\rho_{A|B_{N-1}})]^\alpha
	\end{eqnarray}
	conditioned that
	$E_{\mathcal{\widetilde{N}}}^{4\ln2}(\rho_{A|B_i})\geqslant E_{\mathcal{\widetilde{N}}}^{4\ln2}(\rho_{A|B_{i+1}\cdots B_{N-1}})$ for $i=0,1,\cdots, t$, and $E_{\mathcal{\widetilde{N}}}^{4\ln2}(\rho_{A|B_j})\leqslant E_{\mathcal{\widetilde{N}}}^{4\ln2}(\rho_{A|B_{j+1}\cdots B_{N-1}})$ for $j=t+1,\cdots, N-2$,  $0\leqslant t\leqslant N-3$, $N\geqslant3$.

\noindent{[\bf Proof]}. From Theorem 1 for the case $N=2$, we have
	\begin{eqnarray}\label{thm14:2}
		[E_{\mathcal{\widetilde{N}}}(\rho_{A|B_0\ldots B_{N-1}})]^\alpha
		&\geqslant& [E_{\mathcal{\widetilde{N}}}(\rho_{A|B_0})]^\alpha+\frac{\alpha}{4\ln2} [E_{\mathcal{\widetilde{N}}}(\rho_{A|B_1\ldots B_{N-1}})]^\alpha\nonumber\\
		&\geqslant& \cdots\nonumber\\
		&\geqslant& \sum_{j=0}^{t}(\frac{\alpha}{4\ln2})^{j}[E_{\mathcal{\widetilde{N}}}(\rho_{A|B_{j}})]^\alpha+(\frac{\alpha}{4\ln2})^{t+1}[E_{\mathcal{\widetilde{N}}}(\rho_{A|B_t+1\ldots B_{N-1}})]^\alpha.
	\end{eqnarray}
	Since $E_{\mathcal{\widetilde{N}}}^{4\ln2}(\rho_{A|B_j})\leqslant E_{\mathcal{\widetilde{N}}}^{4\ln2}(\rho_{A|B_{j+1}\cdots B_{N-1}})$ for $j=t+1,\cdots, N-2$, using Theorem 1 again we have
	\begin{eqnarray}\label{thm14:3}
		[E_{\mathcal{\widetilde{N}}}(\rho_{A|B_t+1\ldots B_{N-1}})]^\alpha
		&\geqslant&\frac{\alpha}{4\ln2} [E_{\mathcal{\widetilde{N}}}(\rho_{A|B_{t+1}})]^\alpha+[E_{\mathcal{\widetilde{N}}}(\rho_{A|B_t+2\ldots B_{N-1}})]^\alpha\nonumber\\
		&\geqslant& \cdots\nonumber\\
		&\geqslant& \frac{\alpha}{4\ln2}\left(\sum_{j=t+1}^{N-2}[E_{\mathcal{\widetilde{N}}}(\rho_{A|B_j})]^\alpha\right)+[E_{\mathcal{\widetilde{N}}}(\rho_{A| B_{N-1}})]^\alpha.
	\end{eqnarray}
Combining (\ref{thm14:2}) and (\ref{thm14:3}), we get the inequality (\ref{ger1}).

\noindent{[\bf Remark 3]}.
	From Theorem 3, if $E_{\mathcal{\widetilde{N}}}^{4\ln2}(\rho_{A|B_i})\geqslant E_{\mathcal{\widetilde{N}}}^{4\ln2}(\rho_{A|B_{i+1}\cdots B_{N-1}})$ for all $j=0,1,\cdots, N-2$,  one has
	\begin{eqnarray}\label{thm14:4}
		[E_{\mathcal{\widetilde{N}}}(\rho_{A|B_0B_1\ldots B_{N-1}})]^\alpha
\geq\sum\nolimits_{j=0}^{N-1}\Big(\frac{\alpha}{4\ln2}\Big)^j[E_{\mathcal{\widetilde{N}}}(\rho_{A|B_{j}})]^{\alpha}.
	\end{eqnarray}

For the case of $\alpha<0$, we can also derive a tight upper bound of
$E_{\mathcal{\widetilde{N}}}^{\alpha}(\rho_{A|B_0B_1\ldots B_{N-1}})$.

\noindent{[\bf Theorem 4]}.
For any multiqubit state $\rho_{AB_0\ldots B_{N-1}}$ with $E_{\mathcal{\widetilde{N}}}(\rho_{AB_i})\neq0$,
$i=0,1,\ldots,N-1$,
we have
\begin{equation}\label{SC28}
[E_{\mathcal{\widetilde{N}}}(\rho_{A|B_0B_1\ldots B_{N-1}})]^\alpha
\leq\frac{1}{N}\sum\nolimits_{j=0}^{N-1}[E_{\mathcal{\widetilde{N}}}(\rho_{A|B_j})]^\alpha
\end{equation}
for all $\alpha<0$.

\noindent{[\bf Proof]}.
Similar to the proof in \cite{Fei2}, for arbitrary three-qubit states we have
\begin{align}\label{SCREN1}
[E_{\mathcal{\widetilde{N}}}(\rho_{A|B_0B_1})]^\alpha
\leq & [E_{\mathcal{\widetilde{N}}}^{4\ln2}(\rho_{A|B_0})+ E_{\mathcal{\widetilde{N}}}^{4\ln2}(\rho_{A|B_1})]^{\frac{\alpha}{4\ln2}}\nonumber\\
=& [E_{\mathcal{\widetilde{N}}}(\rho_{A|B_0})]^\alpha \Big(1+\frac{E_{\mathcal{\widetilde{N}}}^{4\ln2}(\rho_{A|B_1})}{E_{\mathcal{\widetilde{N}}}^{4\ln2}(\rho_{A|B_0})}\Big)^{\frac{\alpha}{4\ln2}}\nonumber\\
<& [E_{\mathcal{\widetilde{N}}}(\rho_{A|B_0})]^\alpha,
\end{align}
where the first inequality is from $\alpha<0$, the second inequality is due to
$\Big(1+\frac{E_{\mathcal{\widetilde{N}}}^{4\ln2}(\rho_{A|B_1})}{E_{\mathcal{\widetilde{N}}}^{4\ln2}(\rho_{A|B_0})}\Big)^\frac{\alpha}{4\ln2}<1$.
Moreover, we have
\begin{equation}\label{SCREN2}
[E_{\mathcal{\widetilde{N}}}(\rho_{A|B_0B_1})]^\alpha<[E_{\mathcal{\widetilde{N}}}(\rho_{A|B_1})]^\alpha.
\end{equation}
Combining \eqref{SCREN1} and \eqref{SCREN2}, we get
\begin{equation*}
[E_{\mathcal{\widetilde{N}}}(\rho_{A|B_0B_1})]^\alpha
<\frac{1}{2}\{[E_{\mathcal{\widetilde{N}}}(\rho_{A|B_0})]^\alpha+[E_{\mathcal{\widetilde{N}}}(\rho_{A|B_1})]^\alpha\}.
\end{equation*}
Thus, we obtain
\begin{align}\label{S23}
&[E_{\mathcal{\widetilde{N}}}(\rho_{A|B_0B_1\ldots B_{N-1}})]^\alpha \nonumber\\
< & \frac{1}{2}\Bigg\{\Big[E_{\mathcal{\widetilde{N}}}(\rho_{A|B_0})\Big]^\alpha+\Big[E_{\mathcal{\widetilde{N}}}(\rho_{A|B_1\ldots B_{N-1}})\Big]^\alpha\Bigg\}\nonumber\\
<& \frac{1}{2}\Big[E_{\mathcal{\widetilde{N}}}(\rho_{A|B_0})\Big]^\alpha+\Big(\frac{1}{2}\Big)^2\Big[E_{\mathcal{\widetilde{N}}}(\rho_{A|B_1})\Big]^\alpha
+\Big(\frac{1}{2}\Big)^2\Big[E_{\mathcal{\widetilde{N}}}(\rho_{A|B_2\ldots B_{N-1}})\Big]^\alpha \nonumber\\
<& \ldots \nonumber\\
<& \frac{1}{2}\Big[E_{\mathcal{\widetilde{N}}}(\rho_{A|B_0})\Big]^\alpha+\Big(\frac{1}{2}\Big)^2\Big[E_{\mathcal{\widetilde{N}}}(\rho_{A|B_1})\Big]^\alpha+\ldots
+\Big(\frac{1}{2}\Big)^{N-1}\Big[E_{\mathcal{\widetilde{N}}}(\rho_{A|B_{N-2}})\Big]^\alpha
+\Big(\frac{1}{2}\Big)^{N-1}\Big[E_{\mathcal{\widetilde{N}}}(\rho_{A|B_{N-1}})\Big]^\alpha.
\end{align}
One can get a set of inequalities through the cyclic permutation of the pair indices $B_0$, $B_1,$ $\ldots$, $B_{N-1}$ in \eqref {S23}. Summing up these inequalities, we get \eqref {SC28}.
\qed

\noindent{[\bf Remark4]}.
In (\ref{SC28}) we have assumed that all $E_{\mathcal{\widetilde{N}}}(\rho_{AB_i})$, $i=0,1,2,\cdots,N-1$, are nonzero.
In fact, if one of them is zero, the inequality still holds if one removes this term from the inequality. Namely, if $E_{\mathcal{\widetilde{N}}}(\rho_{AB_i})=0,$ then one has $E_{\mathcal{\widetilde{N}}}^{\alpha}(\rho_{A|B_0B_1\cdots B_{N-1}})<\frac{1}{2}E_{\mathcal{\widetilde{N}}}^{\alpha}(\rho_{A|B_0})+\cdots+\left(\frac{1}{2}\right)^{i}E_{\mathcal{\widetilde{N}}}^{\alpha}(\rho_{A|B_{i-1}})
+\left(\frac{1}{2}\right)^{i+1}E_{\mathcal{\widetilde{N}}}^{\alpha}(\rho_{A|B_{i+1}})+\cdots+\left(\frac{1}{2}\right)^{N-2}E_{\mathcal{\widetilde{N}}}^{\alpha}(\rho_{A|B_{N-2}})
+\left(\frac{1}{2}\right)^{N-2} E_{\mathcal{\widetilde{N}}}^{\alpha}(\rho_{A|B_{N-1}})$. Similar to the analysis in proving Theorem 4, one gets $E_{\mathcal{\widetilde{N}}}^\alpha(\rho_{A|B_0B_1\cdots B_{N-1}})<\frac{1}{N-1}
(E_{\mathcal{\widetilde{N}}}^\alpha(\rho_{A|B_0}+\cdots+E_{\mathcal{\widetilde{N}}}^\alpha(\rho_{A|B_{i-1}})
+E_{\mathcal{\widetilde{N}}}^\alpha(\rho_{A|B_{i+1}})+\cdots+E_{\mathcal{\widetilde{N}}}^\alpha(\rho_{A|B_{N-1}})),
$ for $\alpha<0$.

\section{Tight constraints of multi-qubit entanglement in terms of LCRENoA}
In this section, we provide a class of tight polygamy inequalities of multi-qubit entanglement in
terms of the $\alpha$th-powered LCRENoA and the Hamming weight of the binary vector related to the
distribution of subsystems for $0\leq\alpha\leq2$. For the case of $\alpha<0$, we also propose a monogamy
relation for LCRENoA.

\noindent{[\bf Theorem 5]}. For any multi-qubit state $\rho_{AB_0\ldots B_{N-1}}$ and $0\leq\alpha\leq2$, we have
\begin{equation}\label{LC1}
[E_{\mathcal{\widetilde{N}}_{a}}(\rho_{A|B_0B_1\ldots B_{N-1}})]^\alpha
\leq\sum\limits_{j=0}^{N-1}\Big(\frac{\alpha}{2}\Big)^{\omega_H(\vec{j})}[E_{\mathcal{\widetilde{N}}_{a}}(\rho_{A|B_j})]^\alpha,
\end{equation}
where $0\leq\alpha\leq2$, $\overrightarrow{j}=\left(j_0, \cdots ,j_{n-1}\right)$ is the vector from the binary representation of $j$, and
$\omega_{H}\left(\overrightarrow{j}\right)$ is the Hamming weight of $\overrightarrow{j}$.

\noindent{[\bf Proof]}. From inequality \ref{poly},  one has $E_{\mathcal{\widetilde{N}}_{a}}^{2}(\rho_{A|B_{0}\cdots B_{N-1}})\leq \sum\limits_{i=0}^{N-1}  E_{\mathcal{\widetilde{N}}_{a}}^{2}(\rho_{A|B_{i}})$, thus, it is sufficient to show that
\begin{equation}\label{LC2}
\left[\sum\limits_{j=0}^{N-1} E_{\mathcal{\widetilde{N}}_{a}}^{2}(\rho_{A|B_j})\right]^\frac{\alpha}{2}
\leq\sum\limits_{j=0}^{N-1}\Big(\frac{\alpha}{2}\Big)^{\omega_{H}(\vec{j})}
[E_{\mathcal{\widetilde{N}}_{a}}(\rho_{A|B_j})]^\alpha.
\end{equation}
Without loss of generality, we assume that the qubit subsystems $B_0, \ldots, B_{N-1}$ are so labeled such that
\begin{equation}\label{LC3}
E_{\mathcal{\widetilde{N}}_{a}}^{2}(\rho_{A|B_j})\geq E_{\mathcal{\widetilde{N}}_{a}}^{2}(\rho_{A|B_{j+1}})\geq 0
\end{equation}
for $j=0,1,\ldots,N-2$.

We first show that the inequality \eqref{LC2} holds for the case of $N=2^n$.
For $n=1$, let $\rho_{AB_0}$ and $\rho_{AB_1}$ be the two-qubit reduced density matrices
of a three-qubit pure state $\rho_{AB_0 B_1}$. We obtain
\begin{equation}\label{LC4}
[E_{\mathcal{\widetilde{N}}_{a}}^{2}(\rho_{A|B_0})+E_{\mathcal{\widetilde{N}}_{a}}^{2}(\rho_{A|B_1})]^\frac{\alpha}{2}
=[E_{\mathcal{\widetilde{N}}_{a}}(\rho_{A|B_0})]^\alpha \Big(1+\frac{E_{\mathcal{\widetilde{N}}_{a}}^{2}(\rho_{A|B_1})}{E_{\mathcal{\widetilde{N}}_{a}}^{2}(\rho_{A|B_0})}\Big)^\frac{\alpha}{2}.
\end{equation}
Combining \eqref{bu2} and \eqref{LC3}, we have
\begin{equation}\label{LC5}
\Big(1+\frac{E_{\mathcal{\widetilde{N}}_{a}}^{2}(\rho_{A|B_1})}{E_{\mathcal{\widetilde{N}}_{a}}^{2}(\rho_{A|B_0})}\Big)^\frac{\alpha}{2} \leq
1+\displaystyle\frac{\alpha}{2}\Bigg(\frac{E_{\mathcal{\widetilde{N}}_{a}}(\rho_{A|B_1})}{E_{\mathcal{\widetilde{N}}_{a}}(\rho_{A|B_0})}\Bigg)^\alpha.
\end{equation}
From \eqref{LC4} and \eqref{LC5}, we get
\begin{equation*}
[E_{\mathcal{\widetilde{N}}_{a}}^{2}(\rho_{A|B_0})+E_{\mathcal{\widetilde{N}}_{a}}^{2}(\rho_{A|B_1})]^\frac{\alpha}{2}\leq
[E_{\mathcal{\widetilde{N}}_{a}}(\rho_{A|B_0})]^\alpha+\displaystyle\frac{\alpha}{2}[E_{\mathcal{\widetilde{N}}_{a}}(\rho_{A|B_1})]^\alpha.
\end{equation*}
Therefore, the inequality \eqref{LC2} holds for $n=1$.

We assume that the inequality \eqref{LC2} holds for $N=2^{n-1}$ with $n\geq 2$,
and prove the case of $N=2^n$.
For an $(N+1)$-qubit state $\rho_{AB_0B_1 \cdots B_{N-1}}$ with its two-qubit reduced density matrices $\rho_{AB_j}$ with $j=0, \cdots, N-1$,
we have
\begin{align}\label{LC7}
\left(\sum_{j=0}^{N-1}E_{\mathcal{\widetilde{N}}_{a}}^{2}\left(\rho_{A|B_j}\right)\right)^\frac{\alpha}{2}=
\left(\sum_{j=0}^{2^{n-1}-1}E_{\mathcal{\widetilde{N}}_{a}}^{2}\left(\rho_{A|B_j}\right)\right)^\frac{\alpha}{2}
\left(1+\frac{\sum_{j=2^{n-1}}^{2^n-1}E_{\mathcal{\widetilde{N}}_{a}}^{2}\left(\rho_{A|B_j}\right)}
{\sum_{j=0}^{2^{n-1}-1}E_{\mathcal{\widetilde{N}}_{a}}^{2}\left(\rho_{A|B_j}\right)}\right)^\frac{\alpha}{2}.
\end{align}
Because the ordering of subsystems in Inequality \ref{LC3} implies
\begin{equation*}
0\leq\frac{\sum\nolimits_{j=2^{n-1}}^{2^n-1}E_{\mathcal{\widetilde{N}}_{a}}^{2}(\rho_{A|B_j})}{\sum\nolimits_{j=0}^{2^{n-1}-1}
E_{\mathcal{\widetilde{N}}_{a}}^{2}(\rho_{A|B_j})}\leq 1.
\end{equation*}
Thus, Eq.~(\ref{LC7}) and inequality~(\ref{bu2}) lead us to
\begin{equation*}
\left(\sum_{j=0}^{N-1}E_{\mathcal{\widetilde{N}}_{a}}^{2}\left(\rho_{A|B_j}\right)\right)^\frac{\alpha}{2}
\leq \Bigg(\sum\nolimits_{j=0}^{2^{n-1}-1}E_{\mathcal{\widetilde{N}}_{a}}^{2}(\rho_{A|B_j})\Bigg)^\frac{\alpha}{2}
+\displaystyle\frac{\alpha}{2}\Bigg(\sum\nolimits_{j=2^{n-1}}^{2^n-1}E_{\mathcal{\widetilde{N}}_{a}}^{2}(\rho_{A|B_j})\Bigg)^\frac{\alpha}{2}.
\end{equation*}
According to the induction hypothesis, we get
$$
\Bigg(\sum\nolimits_{j=0}^{2^{n-1}-1}E_{\mathcal{\widetilde{N}}_{a}}^{2}(\rho_{A|B_j})\Bigg)^\frac{\alpha}{2} \leq
\sum\nolimits_{j=0}^{2^{n-1}-1}\Big(\frac{\alpha}{2}\Big)^{\omega_H(\vec{j})}
[E_{\mathcal{\widetilde{N}}_{a}}(\rho_{A|B_j})]^\alpha.
$$
By relabeling the subsystems, the induction hypothesis leads to
$$
\Bigg(\sum\nolimits_{j=2^{n-1}}^{2^n-1}E_{\mathcal{\widetilde{N}}_{a}}^{2}(\rho_{A|B_j})\Bigg)^\frac{\alpha}{2} \leq
\sum\nolimits_{j=2^{n-1}}^{2^n-1}\Big(\frac{\alpha}{2}\Big)^{\omega_H(\vec{j})-1}[E_{\mathcal{\widetilde{N}}_{a}}(\rho_{A|B_j})]^\alpha.
$$
Thus, we have
$$
\Bigg(\sum\nolimits_{j=0}^{2^n-1}E_{\mathcal{\widetilde{N}}_{a}}^{2}(\rho_{A|B_j})\Bigg)^\frac{\alpha}{2}\leq
\sum\nolimits_{j=0}^{2^n-1}\Big(\frac{\alpha}{2}\Big)^{\omega_H(\vec{j})}[E_{\mathcal{\widetilde{N}}_{a}}(\rho_{A|B_j})]^\alpha.
$$
Now consider a $(2^n+1)$-qubit state
\begin{equation}\label{LC8}
\Gamma_{AB_0B_1\ldots B_{2^n-1}}=\rho_{AB_0B_1\ldots B_{N-1}}\otimes \sigma_{B_N\ldots B_{2^n-1}},
\end{equation}
which is the tensor product of $\rho_{AB_0B_1\ldots B_{N-1}}$ and an arbitrary $(2^n-N)$-qubit state $\sigma_{B_N\ldots B_{2^n-1}}$. We have
\begin{equation*}
[E_{\mathcal{\widetilde{N}}_{a}}^{2}(\Gamma_{A|B_0B_1\ldots B_{2^n-1}})]^\frac{\alpha}{2}
\leq\sum\nolimits_{j=0}^{2^n-1}\Big(\frac{\alpha}{2}\Big)^{\omega_H(\vec{j})}[E_{\mathcal{\widetilde{N}}_{a}}(\Gamma_{A|B_j})]^\alpha,
\end{equation*}
where $\Gamma_{A|B_j}$ is the two-qubit reduced density matrix of $\Gamma_{AB_0B_1\ldots B_{2^n-1}}$, $j=0,1,\ldots,2^n-1$. Therefore,
\begin{align*}
[E_{\mathcal{\widetilde{N}}_{a}}^{2}(\rho_{A|B_0B_1\ldots B_{N-1}})]^\frac{\alpha}{2} =&[E_{\mathcal{\widetilde{N}}_{a}}^{2}(\Gamma_{A|B_0B_1\ldots B_{2^n-1}})]^\frac{\alpha}{2}\nonumber\\
\leq& \sum\nolimits_{j=0}^{2^n-1}\Big(\frac{\alpha}{2}\Big)^{\omega_H(\vec{j})}[E_{\mathcal{\widetilde{N}}_{a}}(\Gamma_{A|B_j})]^\alpha \nonumber\\
=& \sum\nolimits_{j=0}^{N-1}\Big(\frac{\alpha}{2}\Big)^{\omega_H(\vec{j})}[E_{\mathcal{\widetilde{N}}_{a}}(\rho_{A|B_j})]^\alpha,
\end{align*}
where $\Gamma_{A|B_0B_1\ldots B_{2^n-1}}$ is separated to the bipartition $AB_0\ldots B_{N-1}$ and $B_N\ldots B_{2^n-1}$, $E_{\mathcal{\widetilde{N}}_{a}}\left(\Gamma_{A|B_0 B_1 \cdots B_{2^n-1}}\right)=E_{\mathcal{\widetilde{N}}_{a}}\left(\rho_{A|B_0 B_1 \cdots B_{N-1}}\right)$,
$E_{\mathcal{\widetilde{N}}_{a}}\left(\Gamma_{A|B_j}\right)=0$ for $j=N, \cdots , 2^n-1$,
and $\Gamma_{AB_j}=\rho_{AB_j}$ for each $j=0, \cdots , N-1$, and this completes the proof.
\qed

\noindent{[\bf Remark 5]}. Since $(\frac{\alpha}{2})^{\omega_H(\overrightarrow{j})}\leq 1$ for  any $0\leq\alpha\leq2$, for any multiqubit state $\rho_{AB_0 B_1 \cdots B_{N-1}}$ we have the following relation
\begin{equation*}
[E_{\mathcal{\widetilde{N}}_{a}}(\rho_{A|B_0B_1\ldots B_{N-1}})]^\alpha
\leq\sum\nolimits_{j=0}^{N-1}\Big(\frac{\alpha}{2}\Big)^{\omega_H(\vec{j})}[E_{\mathcal{\widetilde{N}}_{a}}(\rho_{A|B_j})]^\alpha
\leq\sum\nolimits_{j=0}^{N-1}[E_{\mathcal{\widetilde{N}}_{a}}(\rho_{A|B_j})]^\alpha.
\end{equation*}
Therefore, our inequality \eqref{LC1} in Theorem 5 is always tighter than the inequality \eqref{poly} in Ref.~\cite{LMG}.

\begin{figure}[h]
	\centering
	\scalebox{2.0}{\includegraphics[width=3.9cm]{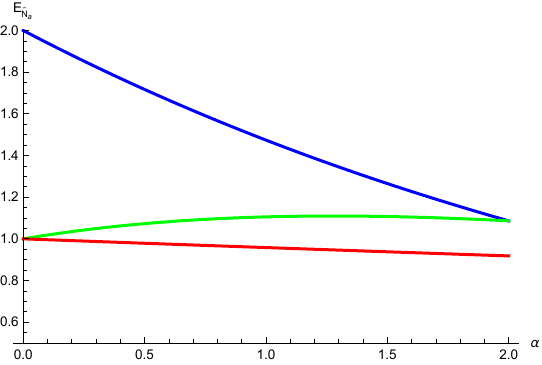}}
	\caption{\small The vertical axis is the the upper bound of the  LCRENOA $E_{\mathcal{\widetilde{N}}_{a}}(\rho_{A|BC})$. The red line is the exact values of  $E_{\mathcal{\widetilde{N}}_{a}}(\rho_{A|BC})$. The  green  line represents the upper bound from our results. The blue line represents the upper bound from the result in \cite{LMG}.}
	\label{Fig3}
\end{figure}

\noindent{[\bf Example 3]}. Let us consider the 3-qubit generalized $W$ state $$|W\rangle_{ABC}=\frac{1}{\sqrt{3}}(|100\rangle+|010\rangle+|001\rangle).$$
We have $\mathcal{\widetilde{N}}_{a}(\rho_{A|BC})={2\sqrt{2}}/{3}$, $\mathcal{\widetilde{N}}_{a}(\rho_{A|B})={2}/{3}$ and $\mathcal{\widetilde{N}}_{a}(\rho_{A|C})={2}/{3}$. Using (\ref{10}) we have $E_{\mathcal{\widetilde{N}}_{a}}(\rho_{A|BC})=\log_{2}({2\sqrt{2}}/{3}+1)$, $E_{\mathcal{\widetilde{N}}_{a}}(\rho_{A|B})=\log_{2}\frac{5}{3}$ and $E_{\mathcal{\widetilde{N}}_{a}}(\rho_{A|C})=\log_{2}\frac{5}{3}$.
Thus, $[E_{\mathcal{\widetilde{N}}_{a}}(\rho_{A|BC})]^\alpha\leq(\log_{2}\frac{5}{3})^\alpha
+\frac{\alpha}{2}(\log_{2}\frac{5}{3})^\alpha$ from our result (\ref{LC1}), and  $[E_{\mathcal{\widetilde{N}}_{a}}(\rho_{A|BC})]^\alpha\leq(\log_{2}(\log_{2}\frac{5}{3})^\alpha
+(\log_{2}\frac{5}{3})^\alpha$ from the result given in Ref.~\cite{LMG}. One can see that our result is better than the result in Ref.~\cite{LMG} for $0\leq\alpha\leq2$, see Fig. \ref{Fig3}.

Similar to the improvement from the inequality \eqref {n18} to the inequality \eqref{n31},
we can also improve the polygamy inequality in Theorem 6. The proof is similar to the Theorem 2.

\noindent{[\bf Theorem 6]}.
For $0\leq\alpha\leq2$, any multiqubit state $\rho_{AB_0\ldots B_{N-1}}$ satisfies
\begin{equation}\label{LC9}
[E_{\mathcal{\widetilde{N}}_{a}}(\rho_{A|B_0B_1\ldots B_{N-1}})]^\alpha
\leq\sum\nolimits_{j=0}^{N-1}\Big(\frac{\alpha}{2}\Big)^j[E_{\mathcal{\widetilde{N}}_{a}}(\rho_{A|B_{j}})]^{\alpha},
\end{equation}
if
\begin{equation}\label{LC10}
E_{\mathcal{\widetilde{N}}_{a}}^{2}(\rho_{A|B_i})\geq\sum\nolimits_{j=i+1}^{N-1}E_{\mathcal{\widetilde{N}}_{a}}^{2}(\rho_{A|B_j})
\end{equation}
for $i=0,1,\ldots,N-2$.

\noindent{[\bf Remark 6]}. In fact, according to \eqref {n11}, for any $0\leq\alpha\leq2$, one has
\begin{align*}
[E_{\mathcal{\widetilde{N}}_{a}}(\rho_{A|B_0\ldots B_{N-1}})]^\alpha
\leq & \sum_{j=0}^{N-1} \left( \frac{\alpha}{2}\right)^{j}\left(E_{\mathcal{\widetilde{N}}_{a}}\left(\rho_{A|B_j}\right)\right)^{\alpha} \nonumber\\
\leq & \sum\nolimits_{j=0}^{N-1}\Big(\frac{\alpha}{2}\Big)^{\omega_H(\vec{j})}[E_{\mathcal{\widetilde{N}}_{a}}(\rho_{A|B_j})]^\alpha.
\end{align*}
Therefore the inequality (\ref{LC9}) of Theorem 6 is tighter than  the inequality (\ref{LC1}) of Theorem 5 under certain conditions.

 We can also provide a more general result by changing the conditions of the Theorem 6. The proof is similar to the Theorem 3.

\noindent{[\bf Theorem 7]}.
For $0\leq\alpha\leq2$, any multiqubit state $\rho_{AB_0\ldots B_{N-1}}$ satisfies
\begin{eqnarray}\label{ger2}
		[E_{\mathcal{\widetilde{N}}_{a}}(\rho_{A|B_0\ldots B_{N-1}})]^\alpha&\leq& \sum_{j=0}^{t}(\frac{\alpha}{2})^{j}[E_{\mathcal{\widetilde{N}}_{a}}(\rho_{A|B_{j}})]^\alpha+(\frac{\alpha}{2} )^{t+2}\sum_{j=t+1}^{N-2}[E_{\mathcal{\widetilde{N}}_{a}}(\rho_{A|B_{j}})]^\alpha\nonumber\\
		&&\ \ \
		+(\frac{\alpha}{2})^{t+1}[E_{\mathcal{\widetilde{N}}_{a}}(\rho_{A|B_{N-1}})]^\alpha
	\end{eqnarray}
	conditioned that
	$E_{\mathcal{\widetilde{N}}_{a}}^{2}(\rho_{A|B_i})\geqslant E_{\mathcal{\widetilde{N}}_{a}}^{2}(\rho_{A|B_{i+1}\cdots B_{N-1}})$ for $i=0,1,\cdots, t$ and $E_{\mathcal{\widetilde{N}}_{a}}^{2}(\rho_{A|B_j})\leqslant E_{\mathcal{\widetilde{N}}_{a}}^{2}(\rho_{A|B_{j+1}\cdots B_{N-1}})$ for $j=t+1,\cdots, N-2$,  $0\leqslant t\leqslant N-3$, $N\geqslant3$.

\noindent{[\bf Remark 7]}.
	From Theorem 7, if $E_{\mathcal{\widetilde{N}}_{a}}^{2}(\rho_{AB_i})\geqslant E_{\mathcal{\widetilde{N}}_{a}}^{2}(\rho_{A|B_{i+1}\cdots B_{N-1}})$ for all $j=0,1,\cdots, N-2$,  one has
	\begin{eqnarray}
		[E_{\mathcal{\widetilde{N}}_{a}}(\rho_{A|B_0B_1\ldots B_{N-1}})]^\alpha
\leq\sum\nolimits_{j=0}^{N-1}\Big(\frac{\alpha}{2}\Big)^j[E_{\mathcal{\widetilde{N}}_{a}}(\rho_{A|B_{j}})]^{\alpha}.
	\end{eqnarray}

For the case of $\alpha<0$, similar to the  Theorem 4, we can also derive a tight lower bound of
$E_{\mathcal{\widetilde{N}}_{a}}^{\alpha}(\rho_{A|B_0B_1\ldots B_{N-1}})$.

\noindent{[\bf Theorem 8]}.
For any multiqubit state $\rho_{AB_0\ldots B_{N-1}}$ with $E_{\mathcal{\widetilde{N}}_{a}}(\rho_{AB_i})\neq0$,
$i=0,1,\ldots,N-1$,
we have
\begin{equation}\label{LC12}
[E_{\mathcal{\widetilde{N}}_{a}}(\rho_{A|B_0B_1\ldots B_{N-1}})]^\alpha
\geq\frac{1}{N}\sum\nolimits_{j=0}^{N-1}[E_{\mathcal{\widetilde{N}}_{a}}(\rho_{A|B_j})]^\alpha,
\end{equation}
for all $\alpha<0$.

\noindent{[\bf Remark 8]}.
In (\ref{LC12}) we have assumed that all $E_{\mathcal{\widetilde{N}}_{a}}(\rho_{AB_i})$, $i=0,1,2,\cdots,N-1$, are nonzero.
In fact, if one of them is zero, the inequality still holds if one removes this term from the inequality. Namely, if $E_{\mathcal{\widetilde{N}}_{a}}(\rho_{AB_i})=0,$ then one has $E_{\mathcal{\widetilde{N}}_{a}}^{\alpha}(\rho_{A|B_0B_1\cdots B_{N-1}})\geq\frac{1}{2}E_{\mathcal{\widetilde{N}}_{a}}^{\alpha}(\rho_{A|B_0})+\cdots+\left(\frac{1}{2}\right)^{i}E_{\mathcal{\widetilde{N}}_{a}}^{\alpha}(\rho_{A|B_{i-1}})
+\left(\frac{1}{2}\right)^{i+1}E_{\mathcal{\widetilde{N}}_{a}}^{\alpha}(\rho_{A|B_{i+1}})+\cdots+\left(\frac{1}{2}\right)^{N-2}E_{\mathcal{\widetilde{N}}_{a}}^{\alpha}(\rho_{A|B_{N-2}})
+\left(\frac{1}{2}\right)^{N-2} E_{\mathcal{\widetilde{N}}_{a}}^{\alpha}(\rho_{A|B_{N-1}})$. Similar to the analysis in proving Theorem 4, one gets $E_{\mathcal{\widetilde{N}}_{a}}^\alpha(\rho_{A|B_0B_1\cdots B_{N-1}})\geq\frac{1}{N-1}
(E_{\mathcal{\widetilde{N}}_{a}}^\alpha(\rho_{A|B_0}+\cdots+E_{\mathcal{\widetilde{N}}_{a}}^\alpha(\rho_{A|B_{i-1}})
+E_{\mathcal{\widetilde{N}}_{a}}^\alpha(\rho_{A|B_{i+1}})+\cdots+E_{\mathcal{\widetilde{N}}_{a}}^\alpha(\rho_{A|B_{N-1}})),
$ for $\alpha<0$.

\section{TIGHT MONOGAMY CONSTRAINT OF MULTIPARTY ENTANGLEMENT BEYOND QUBITS}
Actually, the tight monogamy inequality (\ref{n18}) is applicable not just to multi-qubit systems, but also to certain multipartite higher-dimensional quantum systems. In this section, we will show our monogamy inequality is  more effective for the counterexamples of the CKW monogamy inequality in higher-dimensional systems. We first recall the conceptions of  tangle and it's multi-qubit monogamy  inequalities. The tangle of a bipartite pure states $|\psi\rangle_{AB}$ is defined as \cite{CKW}
\begin{equation}
\tau(|\psi\rangle_{A|B})=2(1-{\rm tr}\rho_A^2),
\end{equation}
where $\rho_A={\rm tr}_B|\psi\rangle_{AB}\langle\psi|$.
The tangle of a bipartite mixed state $\rho_{AB}$ is defined as
\begin{equation}\label{tauAB}
\tau(\rho_{A|B})=\Bigg[\min\limits_{\{p_k,|\psi_k\rangle\}}\sum\limits_{k}p_k\sqrt{\tau(|\psi_k\rangle_{A|B})}\Bigg]^2,
\end{equation}
where the minimization in \eqref{tauAB} is taken over all possible pure state decompositions of $\rho_{AB}=\sum\nolimits_{k}p_k|\psi_k\rangle_{AB}\langle\psi_k|$.

By using tangle, the monogamy inequality of multiqubit
entanglement was proposed as
\begin{equation}\label{h1}
\mathcal\tau(\rho_{A|B_0B_{1}\cdots A_N-1})\geq\sum\limits_{j=0}^{N-1}\tau(\rho_{A|B_j}),
\end{equation}
where $\tau(\rho_{A|B_{0}\cdots B_{N-1}})$ is the tangle of $\rho_{A B_0 \cdots B_N-1}$ with
respect to the bipartition between $A$ and $B_0 \cdots B_{N-1}$, and $\tau(\rho_{A|B_{i}})$ is the tangle of the reduced density matrix $\rho_{A B_i}$, $i=0,\cdots,N-1$.

Although the tangle-based monogamy inequality in (\ref{h1}) characterizes the mutually exclusive nature of two-qubit entanglement shared in multi-qubit systems, it it also known to fail in generalization for systems where at least one local dimension is larger than two \cite{KDS, YCO, KJP}; there exists a quantum state in three-qutrit systems (that is, $3\otimes3\otimes3$ quantum system)
\begin{eqnarray}\label{h2}
|\Psi\rangle_{A|BC}&=&\frac{1}{\sqrt{6}}(|012\rangle-|021\rangle+|120\rangle-|102\rangle+|201\rangle-|210\rangle),
\end{eqnarray}
where $\tau(|\Psi\rangle_{ABC})=\frac{4}{3}$ and $\tau(\rho_{AB})=\tau(\rho_{AC})=1$, thus
\begin{eqnarray}\label{h3}
\tau(|\Psi\rangle_{A|BC})<\tau(\rho_{AB})+\tau(\rho_{AC}),
\end{eqnarray}
which implies the violation of inequality (\ref{h1}).

In fact, there exists a quantum state in $3\otimes2\otimes2$ quantum systems \cite{KDS,KJP},
\begin{eqnarray}\label{h4}
|\Psi\rangle_{ABC}&=&\frac{1}{\sqrt{6}}(\sqrt{2}|010\rangle+\sqrt{2}|101\rangle+|200\rangle+|211\rangle,
\end{eqnarray}
where $\tau(|\Psi\rangle_{A|BC})=\frac{4}{3}$ and $\tau(\rho_{AB})=\tau(\rho_{AC})=\frac{8}{9}$, and this
also implies the violation of inequality (\ref{h1}). In other words, tangle-based monogamy inequality in (\ref{h1}) only holds for multi-qubit systems, and even tiny extension in any of the subsystems leads to a violation.

\begin{figure}[h]
	\centering
	\scalebox{2.0}{\includegraphics[width=3.9cm]{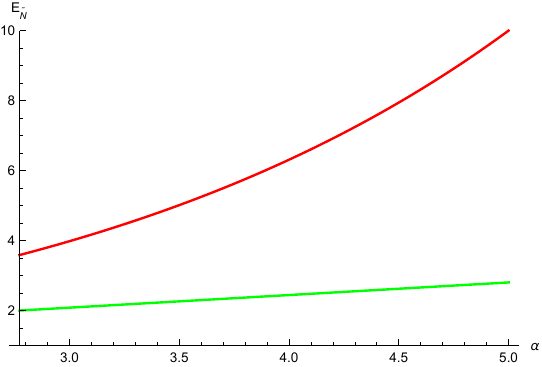}}
	\caption{\small The vertical axis is the the lower bound of the  LCREN $E_{\mathcal{\widetilde{N}}}(|\Psi\rangle_{A|BC})$. The red line is the exact values of  $E_{\mathcal{\widetilde{N}}}(|\Psi\rangle_{A|BC})$. The  green  line represents the lower bound from our results (\ref{n18}).}
	\label{Fig4}
\end{figure}

Let us now consider the LCREN-based monogamy inequality in (\ref{n18}) for the quantum state in Eq.(\ref{h2}). After a bit of calculations, it is straightforward to verify that $\mathcal{\widetilde{N}}(|\Psi\rangle_{A|BC})=2$ and $\mathcal{\widetilde{N}}(\rho_{AB})=\mathcal{\widetilde{N}}(\rho_{AB})=1$. Using (\ref{9}) we have $E_{\mathcal{\widetilde{N}}}(|\Psi\rangle_{A|BC})=\log_{2}3$, $E_{\mathcal{\widetilde{N}}}(\rho_{AB})=E_{\mathcal{\widetilde{N}}}(\rho_{AC})=1$.   Thus, we have
\begin{eqnarray}\label{h5}
E_{\mathcal{\widetilde{N}}}^{\alpha}(|\Psi\rangle_{A|BC})=(\log_{2}3^{\alpha})\geq 1+ \frac{\alpha}{4\ln2}= E_{\mathcal{\widetilde{N}}}^{\alpha}(\rho_{AB})+\frac{\alpha}{4\ln2}E_{\mathcal{\widetilde{N}}}^{\alpha}(\rho_{AC})
\end{eqnarray}
for $\alpha\geq4\ln2$, see Fig. \ref{Fig4}.

Similarly, for the quantum state in Eq.(\ref{h4}), we have $\mathcal{\widetilde{N}}(|\Psi\rangle_{A|BC})=2$ and $\mathcal{\widetilde{N}}(\rho_{AB})=\mathcal{\widetilde{N}}(\rho_{AB})=\frac{2\sqrt{2}}{3}$. Using (\ref{9}) we have $E_{\mathcal{\widetilde{N}}}(|\Psi\rangle_{A|BC})=\log_{2}3$, $E_{\mathcal{\widetilde{N}}}(\rho_{AB})=E_{\mathcal{\widetilde{N}}}(\rho_{AC})=\log_{2}(\frac{2\sqrt{2}}{3}+1)$.   Thus, we have
\begin{eqnarray}\label{h6}
E_{\mathcal{\widetilde{N}}}^{\alpha}(|\Psi\rangle_{A|BC})=(\log_{2}3^{\alpha})\geq (1+ \frac{\alpha}{4\ln2})(\log_{2}(\frac{2\sqrt{2}}{3}+1)) ^{\alpha}= E_{\mathcal{\widetilde{N}}}^{\alpha}(\rho_{AB})+\frac{\alpha}{4\ln2}E_{\mathcal{\widetilde{N}}}^{\alpha}(\rho_{AC})
\end{eqnarray}
for $\alpha\geq4\ln2$.

In other words, the LCREN-based monogamy inequality in (\ref{n18}) is still valid for the counterexamples of tangle-based monogamy inequality. Thus LCREN is a good alternative for monogamy inequality of multi-qubit entanglement even in higher-dimensional quantum systems so far.

\section{Conclusions}
The monogamy and polygamy relations of quantum entanglement are fundamental properties exhibited by multipartite entangled states. We have provided a characterization of multi-qubit entanglement monogamy and polygamy constraints in terms of nonconvex entanglement measures. Using the Hamming weight of the binary vector related with the distribution of subsystems, we have established a class of tight monogamy inequalities of multi-qubit entanglement based on the $\alpha$th-power of LCREN for $\alpha \geq 4\ln2$. We have further established  a class of tight polygamy inequalities of multi-qubit entanglement in terms of the $\alpha$th-power of LCRENoA for $0 \leq \alpha \leq 2$. For the case $\alpha<0$, we give the corresponding polygamy and monogamy relations for LCREN and LCRENoA, respectively. We also show that these new inequalities give rise to tighter constraints than the existing ones. Moreover, our
monogamy inequality is shown to be more effective for the counterexamples of the CKW monogamy inequality in higher-dimensional systems. The entanglement distribution in multipartite systems can be more precisely characterized through stricter monogamy and polygamy inequalities. Our findings may highlight further research on understanding the entanglement distribution in highlight systems.

\end{document}